\documentclass[final,journal,letterpaper]{IEEEtran}

 \usepackage{varioref} 
 \usepackage{wrapfig} 
 \usepackage{threeparttable} 
 \usepackage{dcolumn} 
  \newcolumntype{d}{D{.}{.}{-1}}
 \usepackage{nomencl} 
  \makeglossary
 \usepackage{subfigmat} 
 \usepackage{fancyvrb} 
  \fvset{fontsize=\footnotesize,xleftmargin=2em}
 \usepackage{lettrine} 
 \usepackage[hidelinks]{hyperref}

 \usepackage{epsfig}
 \usepackage{epstopdf}
 \usepackage{subfigure}
 \usepackage{latexsym}
 \usepackage{amsmath}
 \usepackage{amsfonts}
 \usepackage{amssymb}
 \usepackage{mathrsfs}
 \usepackage{cases}
 \usepackage{array}
 \usepackage{color}
  \usepackage{flushend}
   \usepackage{multirow}
 \usepackage{tabularx,booktabs,caption}

 \usepackage{url}
 \usepackage[noadjust]{cite}
 \usepackage{graphicx}
 \usepackage{psfrag}
 \usepackage{color}

\newfont{\Bb}{msbm10 scaled\magstep1}

\begin{document}

\title{Parameters Calibration for Power Grid Stability Models using Deep Learning Methods}

\author{Renke Huang, Rui Fan, Tianzhixi Yin, Shaobu Wang, Zhenyu Tan 
\thanks{Corresponding author: Rui Fan (e-mail: Rui.Fan@pnnl.gov)}
\thanks{R. Huang, R. Fan,  T. Yin, S. Wang are with Pacific Northwest National Laboratory, Richland, WA 99354, USA (e-mail: \{Renke.Huang, Rui.Fan, Tianzhixi.Yin, Shaobu.Wang\}@pnnl.gov).}
\thanks{Z. Tan is with Google, Mountain View, California 94043, USA (e-mail: tanzheny@google.com).}}

\maketitle

\begin{abstract}
This paper presents a novel parameter calibration approach for power system stability models using   automatic data generation and advanced deep learning technology. 
A PMU-measurement-based ``event playback'' approach is used to identify potential inaccurate parameters and automatically generate extensive simulation data, which are used for training a  convolutional neural network (CNN).
The  accurate parameters will be predicted by the well-trained CNN model and validated by original PMU measurements.
The accuracy and effectiveness of the proposed deep learning approach have been validated through extensive simulation  and field data.
\end{abstract}


\section{Introduction}\label{sec:Intro}
\IEEEPARstart{M}{aintaining} high-fidelity power system stability models is of critical importance to ensuring the secure  economic operation and planning of today's power grid considering its increasing stochastic and dynamic behavior.
Traditional power grid stability model validation and parameter calibration are based on stage testing that takes generators offline, which could be quite costly and time-consuming \cite{NERC}.
Over the past few years, the wide  implementation of phasor measurement units (PMUs) has made it possible to directly use online measurements for   model validation.
A particle swarm optimization (PSO)-based approach was proposed for stability model parameter calibration through a simultaneous perturbation stochastic approximation \cite{PSO}. However, this method is time and effort consuming as it requires extensive iterations between dynamic simulations and the PSO.
An advanced ensemble Kalman filter (EnKF)-based method was proposed and integrated with the commercial software package for the generator parameter calibration \cite{EnKF}. However, it requires modification in the source code of the stability models in the commercial software for accommodating the EnKF algorithm; thus, its  accessibility and availability is currently  limited.

In this letter, we have proposed a novel stability model parameter calibration method using   automatic data generation and deep learning technology. An ``event playback'' approach that uses the real and reactive power responses of the under-examining stability model to identify inaccurate parameters is proposed. 
Extensive simulation data will be automatically generated using the   stability model, where the identified parameters are randomly perturbed. 
These generated data are used to train a  convolutional neural network (CNN), in which the output will be the predicted parameter values. The CNN preserves the translation invariance of the input data and it is robust to measurement noises. Advanced deep learning techniques such as rectified linear unit (ReLU) activation and dropout regularization have also been used to improve the neural network performance \cite{Deep}.
Extensive studies using simulated and field data have validated the accuracy and effectiveness of the proposed deep learning approach. 
\vspace{-2mm} 

\section{Proposed Approach}\label{sec:Approach}

\begin{figure}[!ht]
\centering
\includegraphics[width=0.45\textwidth]{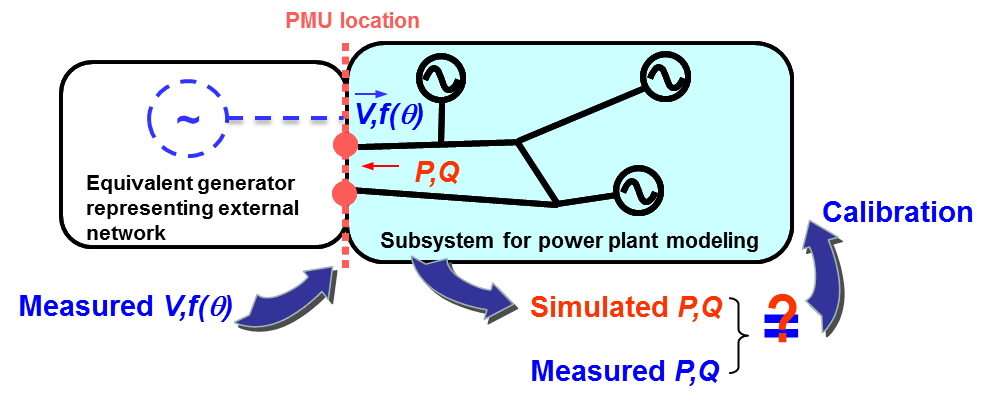}
\vspace{-2mm} 
\caption{The concept of model validation via event playback.}
\label{fig:1}
\end{figure}

The proposed deep learning approach calibrates stability model parameters in a systematic manner. The first step is to determine if inaccurate parameters exist or not  using the ``event playback'' concept shown in Fig. \ref{fig:1}.
The PMU voltage and phase angle  measurements are used to run a   dynamic simulation with acknowledged model parameters. The simulated dynamic response of the subsystem, including active and reactive power, will be compared with the corresponding PMU power measurements. A large difference between the simulated and measured data indicates an discrepancy of the model parameters, thus, a calibration procedure is required.

The second step is to identify a group of potential inaccurate parameters. Because a typical stability model contains dozens of parameters, narrowing the number of targeted parameters would significantly increase the calibration efficiency. A sensitivity analysis method \cite{EnKF} is used to identify potential inaccurate parameters:

\begin{equation}
S(p) = \dfrac{p_0}{2m} \sum_{i=1}^m \left|  \dfrac{r^k(p_1) - r^k(p_2)}{\Delta p_0} \right|
\end{equation}
\textit{s.t.}
\begin{align}
p_1 &= p_0 + \Delta p_0 \\
p_2 &= p_0 - \Delta p_0
\end{align}
where $p_0$ and $\Delta p_0$ are the original value and small perturbation of parameter $p$, $r(\cdot)$ is the time response, and $m$ is the number of total time steps. This sensitivity analysis will wipe out parameters with zero or very small sensitivity.

\begin{figure}[!ht]
\centering
\includegraphics[width=0.46\textwidth]{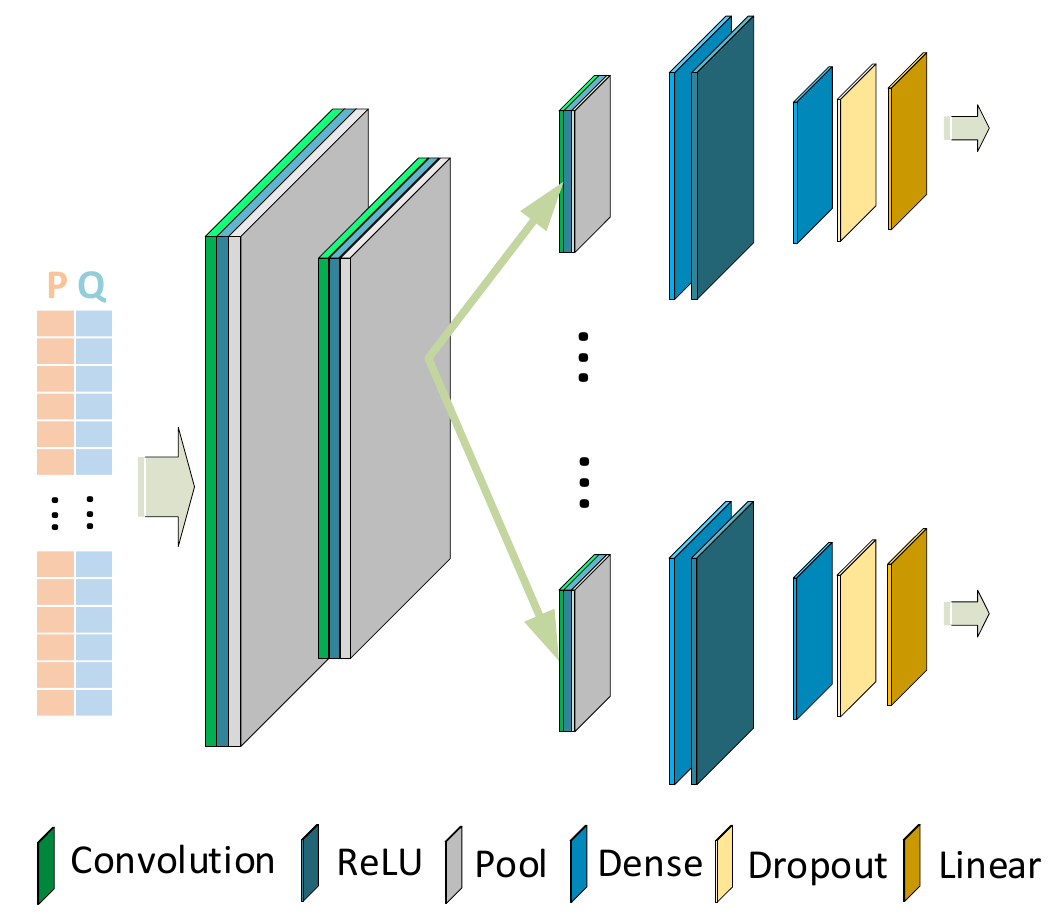}
\caption{Multi-output CNN model for parameter prediction.}
\label{fig:2}
\end{figure}

The third step is to automatically generate extensive generator dynamic data respective to those potential inaccurate parameters. In this paper, we randomly selected parameter values that are 50\% to 200\% of their original values for the data generation. 
The generated data were used to train a multi-output CNN model to predict the parameter values, as shown in Fig. \ref{fig:2}. The first two layers of this CNN model are convolutional layers, and each layer consists of a convolution function, a ReLU activation and a maximum pooling. The output of the convolution and ReLU is

\begin{equation}
z = max\left( 0 , \left(   \sum_i^{N} K[i] \times Y[i]  + B \right)  \right) 
\end{equation}
where $K$ and $Y$ are the filter feature and input,  $N$ is the filter size, and $B$ is the bias term. The maximum pooling will simply take the largest value in the pooling region as the output of the convolutional layer. The objective of the first two convolutional layers is to capture some common detailed input data patterns, such as a sudden drop or increase in the waveform. Parallel branches of convolutional and dense layers are connected to the two common layers for  predicting different parameters individually. The proposed  structure enables feature sharing of  CNN parameters, which dramatically reduces the computation burden.
Dropout regularization layers are used in the CNN model to prevent over-fitting problems. The CNN model uses the root-mean-square (RMS) error  as the loss function for training the neural network.
\vspace{-4mm} 

The final step is feeding the measured PMU data to the well-trained CNN model to  predict the values of those potential inaccurate parameters. The accuracy of the proposed deep learning approach can be validated through another round of ``event playback'' that compares the measured PMU data with the simulated data using the CNN predicted parameters.

\section{Test Results}\label{sec:results}

The developed approach was tested on a real-world power generating unit, which consists of a synchronous generator, an exciter, a power system stabilizer, and a governor. The corresponding standard models are GENROU, ESST1A, GGOV1, and PSS2A \cite{PSSE}.
When large changes occurred in the measured PMU data, we ran the ``event playback'' and noticed a mismatch between the measured and simulated data. 
The mismatch indicated an discrepancy of the model parameters. A sensitivity analysis was carried out and the identified candidate parameters with higher sensitivity are the inertial $H$ for the generator, the gain $K_a$ and time constant $T_b$ for the exciter, and the gain $K_s$ for the governor.

Around 10,000 simulated generator dynamic response data were automatically generated with the targeted parameters that randomly varied between 50\% to 200\% of the original values through ``event playback''. 90\%  data were used for the training and validation of the CNN model, and the other 10\% data were used for testing. We also compared the CNN model with a conventional machine learning neural network called multi-layer perceptron  (MLP).  The two models have similar sizes and numbers of layers, and therefore are comparable.

\emph{Remark 1}: A Windows server with 32 cores of 3.20 GHz Intel Xeon CPUs, 128G Memory, was used to generate the simulation data in parallel, and it took around 30 minutes to generate the 10,000  data. The training of the CNN model took another 15 minutes by using the TensorFlow library \cite{TF}.

\begin{figure}[!ht]
\centering
\includegraphics[width=0.48\textwidth]{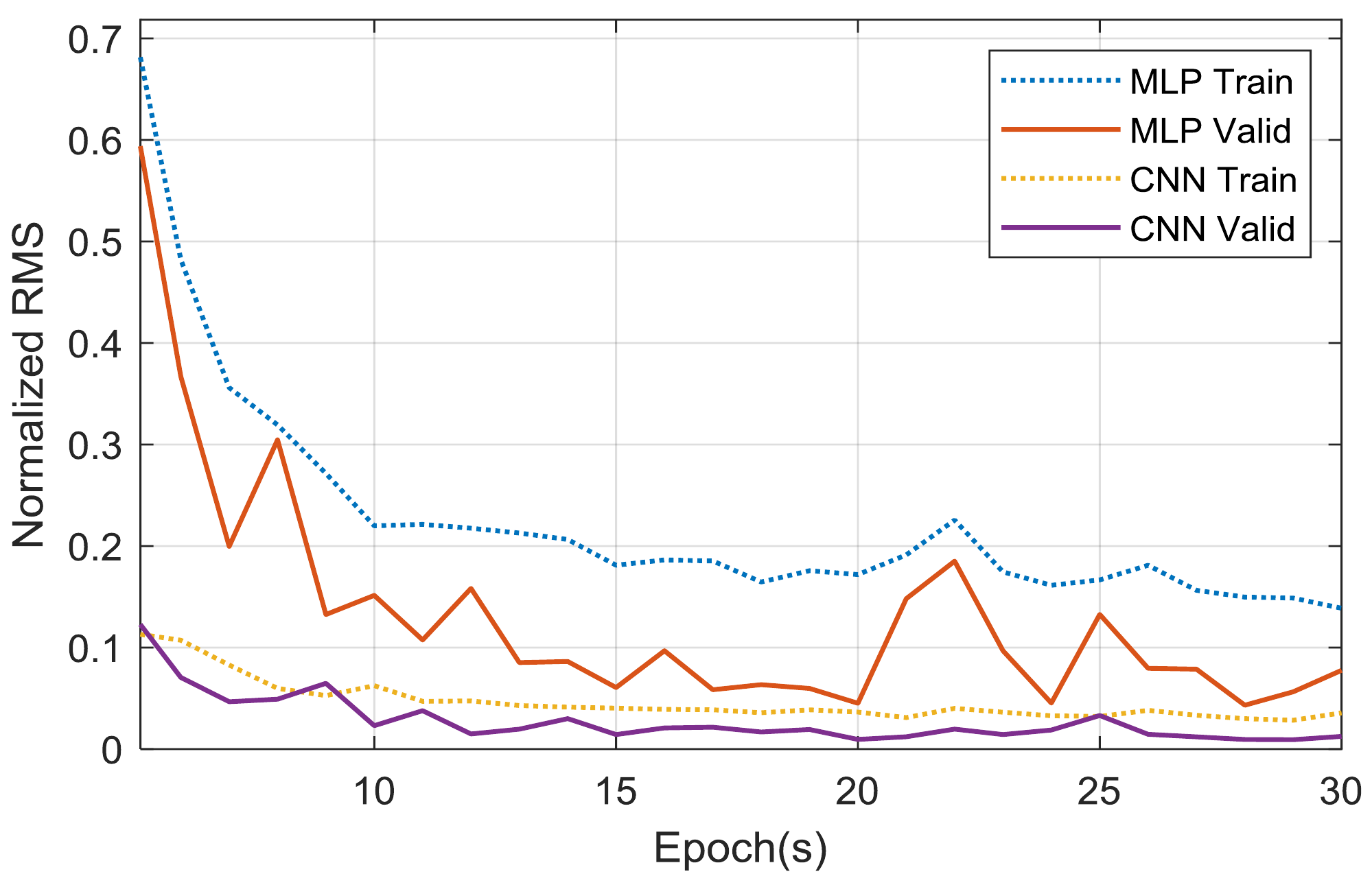}
\caption{Training and validation normalized losses for both the CNN and MLP.}
\label{fig:3}
\end{figure}

\begin{figure}[!ht]
\centering
\vspace{-2mm}
\subfigure[CNN results]{\includegraphics[width=0.47\textwidth]{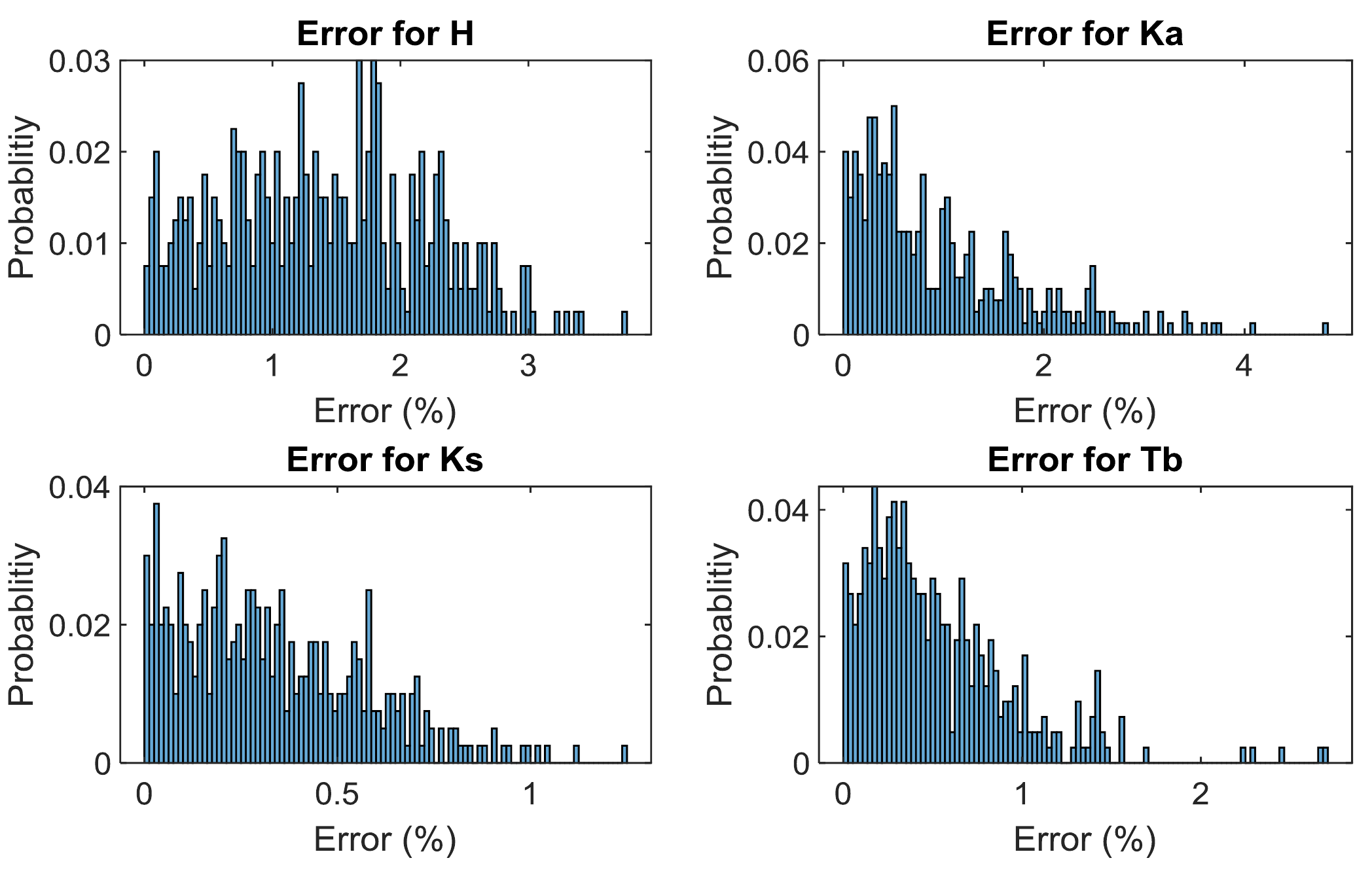}}
\subfigure[MLP results]{\includegraphics[width=0.47\textwidth]{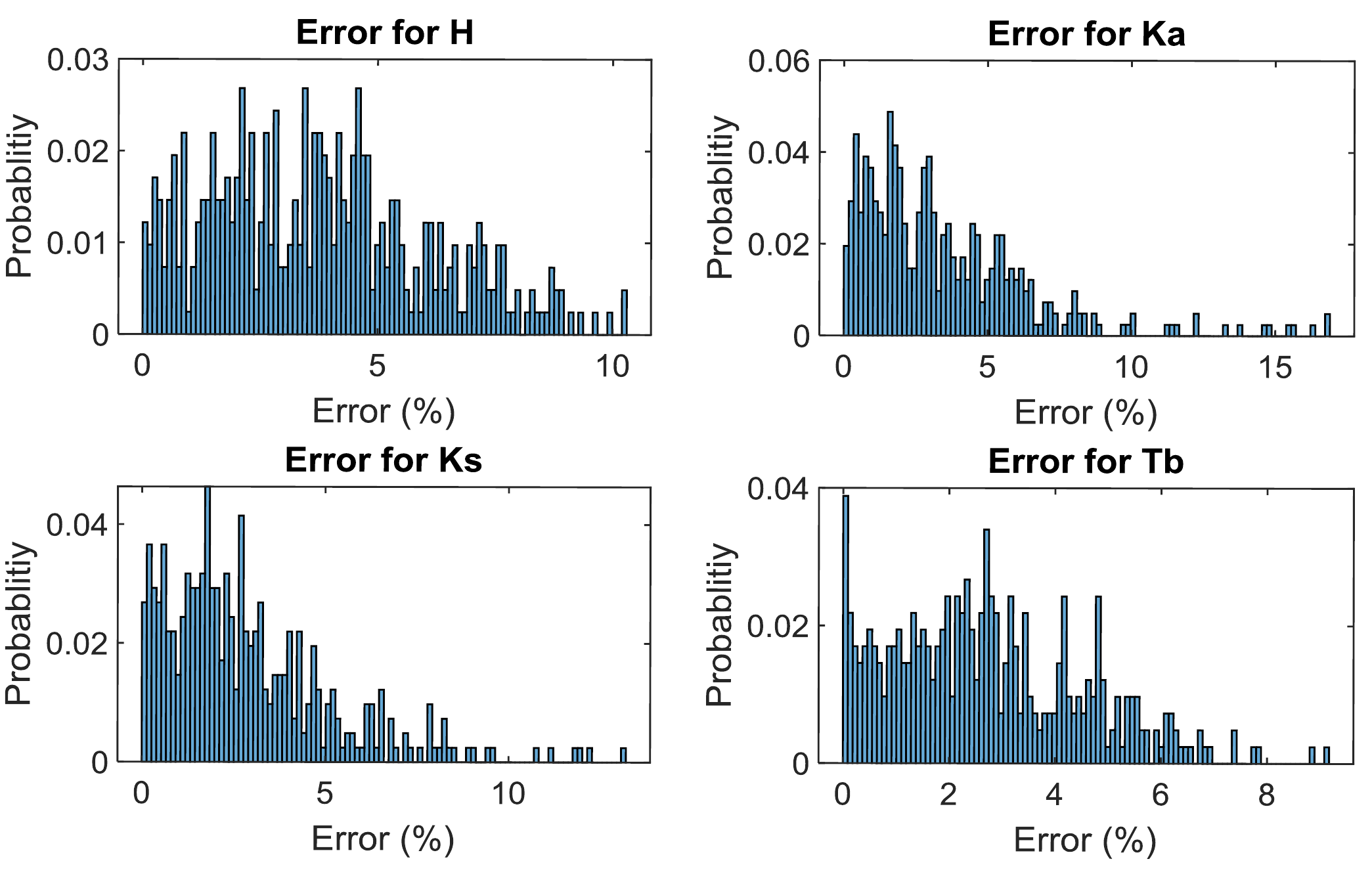}}
\caption{Absolute error histograms of targeted parameters: (a) CNN; (b) MLP.}
\label{fig:4}
\vspace{-2mm} 
\end{figure}

Fig. \ref{fig:3} shows the normalized losses (RMS error) of the training and validation process for both the CNN and MLP models. The CNN model has much smaller losses than the MLP model, indicating the CNN is able to predict the parameter values more accurately. Note that the training loss is higher than the validation loss, because the dropout regularization is used during the training process and deactivated in the validation process. The absolute error histograms of the above targeted parameters in the testing results are shown in Fig. \ref{fig:4}, and the corresponding statistics are presented in Table \ref{table1}.  
It is shown that the prediction errors of the CNN model are much smaller than those of the MLP model.

\begin{table}[htb]
\renewcommand{\arraystretch}{1.2}
\centering
\captionsetup{justification = centering}
\caption{Statistics of the absolute error of targeted parameters}
\begin{tabular}{|c|c|c|c|c|}
\hline
\multirow{ 2}{*}{\textbf{Parameter}} & \multicolumn{2}{c|}{\textbf{MLP Error (\%)}} & \multicolumn{2}{c|}{\textbf{Cnn Error (\%)}}\\
\cline{2-5}
& \textit{Mean} & \textit{Max} & \textit{Mean} & \textit{Max} \\
\hline
$H$ & 3.76  & 10.30 & 1.39 & 3.77 \\
\hline
$K_a$ & 3.46  & 16.84 & 0.96 & 4.83 \\
\hline
$T_b$ & 2.69  & 9.16 & 0.52 & 2.70 \\
\hline
$K_s$ & 2.86  & 13.14 &  0.33 & 1.25 \\
\hline
\end{tabular}
\label{table1}
\end{table}

\begin{figure}[!ht]
\centering
\includegraphics[width=0.48\textwidth]{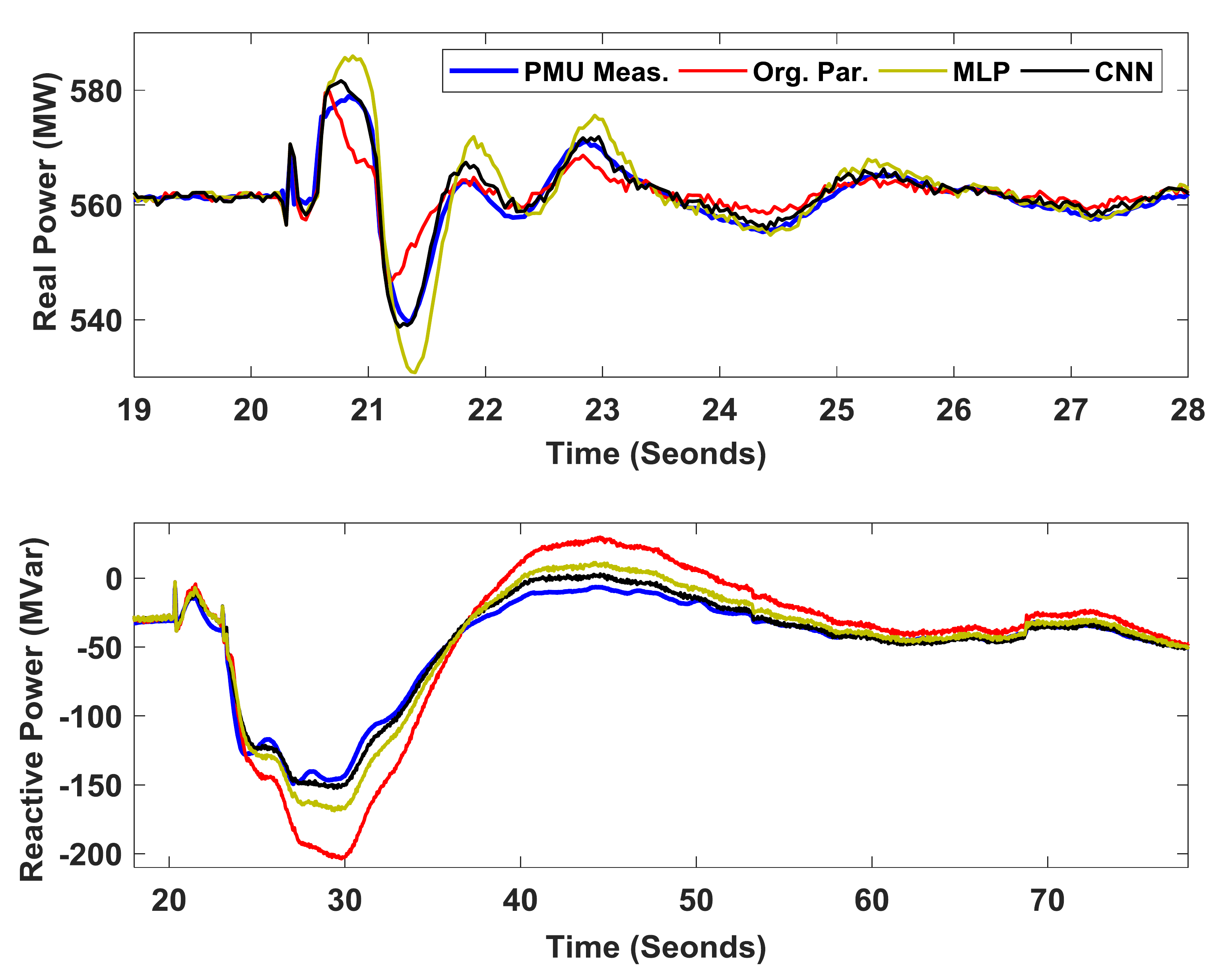}
\caption{Comparison of the real-world PMU data with simulated data.}
\label{fig:5}
\end{figure}

The well-trained CNN and MLP models are used to calibrate the targeted parameters by feeding the actual PMU measurements to the neural network models. The results are listed in Table  \ref{table:2}.
The calibrated parameters of both the CNN and MLP models are obviously different from the original  values, with the exception of the parameter $T_b$ (less than 2\%).
To test the effectiveness of the deep learning approaches, real-world PMU measurements are compared with the simulated data that are generated with the calibrated parameters of both the CNN and MLP models, as shown in Fig. \ref{fig:5}. The PMU measurements (in blue) are largely different than simulated data using original parameters (in red). The  calibration accuracy has been improved after using either the CNN (in black) or the MLP (in brown) method. The CNN method has outperformed the MLP method, as its generated curves fit the PMU measurements better than those of the MLP method.

\begin{table}[!ht]
\renewcommand{\arraystretch}{1.2}
\caption{Parameter calibration results}
\label{table:2}
\centering
\begin{tabular}{c|c|c|c}
\hline
\textbf{Parameter} & \textbf{Original} & \textbf{CNN-Predicted} & \textbf{MLP-Predicted}\\
\hline
$H$ & 4.60 & 5.71 & 5.22 \\
\hline
$K_a$ & 250.00 & 300.97 & 272.45 \\
\hline
$T_b$ & 43.00 & 42.61 & 43.68 \\
\hline
$K_s$ & 36.00 & 30.12 & 32.85 \\
\hline
\end{tabular}
\end{table}

\section{Conclusion}\label{sec:Conc}
This letter proposes a novel parameter calibration approach for power system stability models. We have developed a systematic procedure to identify and calibrate potential inaccurate model parameters, using   automatic data generation and advanced deep learning techniques. 
The performance of the proposed CNN model has been compared with a conventional MLP model using both simulated data and real-world data.
Results have validated the accuracy and effectiveness of the proposed CNN-based approach in calibrating stability model parameters.

\bibliographystyle{IEEEtran}
\bibliography{CNN}

\vfill

\end{document}